\documentclass[9pt,twocolumn,twoside]{pnas-new}
\templatetype{pnasresearcharticle} % Choose template 
\usepackage{ulem}
\title{Two-dimensionally stable self-organization arises in simple schooling swimmers through hydrodynamic interactions}

\author[a,1]{Pedro C. Ormonde}
\author[b,1]{Melike Kurt} 
\author[a]{Amin Mivehchi} 
\author[a,2]{Keith W. Moored}

\affil[a]{Mechanical Engineering and Mechanics, Lehigh University, Bethlehem, PA, 18015, USA}
\affil[b]{Maritime Engineering Group, Faculty of Engineering and the Environment, University of Southampton, Southampton SO17 1BJ, UK}

\leadauthor{Ormonde \& Kurt} 

\significancestatement{Fish schools are fascinating examples of self-organization in nature. They serve many purposes from enhanced foraging, and protection to improved socialization and migration.  However, our understanding of the hydrodynamic interactions in schools is primitive.  It has been postulated that these interactions can regulate energy usage and speed, as well as push and pull individuals thereby altering the school's structure and function.  We have discovered that stable formations of swimmers can arise in two-dimensional schools through passive hydrodynamic forces alone.  In these stable formations, swimmers also experience speed and efficiency benefits.  This opens the door to considering that the structure and function of fish schools may be more strongly regulated by hydrodynamic interactions than previously known.  
}

\authorcontributions{M.K. helped design the study, gathered and processed constrained-foil measurements, drafted and revised the manuscript. P.C.O. helped design the study, gathered and processed the constrained and unconstrained-foil data, gathered the flow-visualization data and helped draft and revise the manuscript. A.M. helped design the study, gathered and processed the numerical and the flow-visualization data, and helped revise the manuscript.  K.W.M. helped design the study, and helped revise the manuscript.}
\authordeclaration{The authors declare no conflict of interest.}
\equalauthors{\textsuperscript{1}P.C.O. contributed equally to this work with M.K.}
\correspondingauthor{\textsuperscript{2}To whom correspondence should be addressed. E-mail: kmoored@lehigh.edu}

\keywords{collective locomotion $|$ hydrodynamic interactions $|$ fish schooling $|$ pattern formation $|$ collective performance} 

\begin{abstract}
We present new constrained and free-swimming experiments and simulations of a pair of pitching hydrofoils interacting in a minimal school. The hydrofoils have an out-of-phase synchronization and they are varied through in-line, staggered, and side-by-side formations within the two-dimensional interaction plane. It is discovered that there is a \textit{two-dimensionally} stable equilibrium point for a side-by-side formation. In fact, this formation is super-stable, meaning that hydrodynamic forces will passively maintain this formation even under external perturbations and the school as a whole has no net forces acting on it that cause it to drift to one side or the other. Moreover, previously discovered \textit{one-dimensionally} stable equilibria driven by wake vortex interactions are shown to be, in fact, two-dimensionally \textit{unstable}, at least for an out-of-phase synchronization. Additionally, it is discovered that a trailing-edge vortex mechanism provides the restorative force to stabilize a side-by-side formation. The stable equilibrium is further verified by experiments and simulations for freely-swimming foils where dynamic recoil motions are present. When constrained, swimmers in compact side-by-side formations experience collective efficiency and thrust increases up to 40\% and 100\%, respectively, whereas slightly staggered formations output an even higher efficiency improvement of 84\% with a 87\% increase in thrust. Freely-swimming foils in a stable side-by-side formation show an efficiency and speed enhancement of up to 9\% and 15\%, respectively. These newfound schooling performance and stability characteristics suggest that fluid-mediated equilibria may play a role in the control strategies of schooling fish and fish-inspired robots.

\end{abstract}

\DeclareUnicodeCharacter{2212}{-}

\begin{document}

\maketitle
\thispagestyle{firststyle}
\ifthenelse{\boolean{shortarticle}}{\ifthenelse{\boolean{singlecolumn}}{\abscontentformatted}{\abscontent}}{}

\dropcap{S}elf-organization of living systems is one of Nature's most ubiquitous and mesmerizing phenomena. It arises across a wide range of spatial and temporal scales from the cells in our bodies \cite{George2017} and swarming of microorganisms \cite{Koch2011} to the flocking of birds \cite{lissaman1970formation} and schooling of fish \cite{weihs1973hydromechanics}. For macroscopic flyers and swimmers, a wide range of hypotheses have attributed collective behavior to social interactions~\cite{wynne1962animal}, protection against predators~\cite{tinbergen2012social}, food prospect optimization~\cite{pitcher1982fish}, and/or energetic benefits \cite{weihs1973hydromechanics}.  Our knowledge of the latter hypothesis is limited since it is regulated by complex hydrodynamic interactions.  Yet, both the spatial organization \cite{lissaman1970formation, badgerow1981energy} and temporal synchronization \cite{drucker2001locomotor, portugal2014upwash, ashraf2017simple} have emerged as factors influencing the hydrodynamic interactions, and, consequently, the energetic cost of locomotion and traveling speed of individuals in a collective.  Still, our understanding of the force production and energetics of schooling swimmers is mostly limited to canonical spatial formations such as a leader-follower in-line formation \cite{Boschitsch2014,muscutt2017performance,kurt2018flow,Heydari2020} and a side-by-side formation \cite{Dewey2014,quinn2014unsteady,ashraf2017simple,Kurt2018aviation}, while there are fewer studies of staggered formations \cite{verma2018efficient,dai2018stable,oza2019lattices,Kurt2020}.  

Because of these studies it is commonly presumed that the spatial organization observed in schools is driven by the interest to maximize swimming efficiency or force production. However, another explanation was first proposed by Sir James Lighthill \cite{lighthill1975mathematical}.  The so-called Lighthill conjecture \cite{Ramananarivo2016} postulates that the formations of fish in a school may be due to the interaction forces that push and pull the swimmers into a particular stable formation, much like the atoms in a crystal lattice.  Indeed, this idea of passive self-organization has shown promise in recent studies where one-dimensional streamwise stability has been observed in pairs of in-line self-propelled foils \cite{becker2015hydrodynamic,Ramananarivo2016} or in small schools of various formations \cite{dai2018stable,Peng2018}, as well as in pairs of in-line hydrofoils with differing kinematics \cite{newbolt2019flow}.  While these studies have shown seminal results supporting the Lighthill conjecture, they have only probed the one-dimensional stability of formations.  However, \textit{two-dimensionally} or even three-dimensionally stable formations are required for the passive self-organization of schools that produce two-dimensional or three-dimensional flows.

Here, we advance our understanding of the hydrodynamic interactions of schooling swimmers in two ways.  For the first time, we measure the \textit{two-dimensional} stability of schooling formations for constrained and freely swimming foils, which takes us closer to understanding the role of the Lighthill conjecture in schooling formations. We discover that many of the one-dimensionally stable formations previously observed are, in fact, unstable once the cross-stream stability is considered.  Yet, we still find that a side-by-side formation is \textit{two-dimensionally} stable, providing support for the hypothesis that this formation observed in real fish \cite{ashraf2017simple} may be due to passive self-organization. Second, we measure the force production and energetics of two interacting hydrofoils throughout a plane of possible formations ranging from in-line to side-by-side by passing through the possible staggered formations.  We reveal that there is a thrust and efficiency optimum in a slightly-staggered formation where there is direct vortex impingement on the follower.   
%%%%%%%%%%%%%%%%%%%%%%%%%%%%

%%%%%%%%%%%%%%%%%%%%%%%%%%%%

\section*{Experimental Approach and Results}
To examine the flow interactions that occur in schools, full swimmer models can be readily used in numerical studies \cite{verma2018efficient}, however, these models are difficult to implement experimentally.  Instead, experiments typically use oscillating hydrofoils as a simple model of the propulsive appendages of animals \cite{Boschitsch2014,Dewey2014,quinn2014unsteady,becker2015hydrodynamic,Ramananarivo2016,muscutt2017performance,kurt2018flow,newbolt2019flow,kurt2019swimming,Kurt2020}.  Importantly, these oscillating hydrofoils capture the salient unsteady fluid mechanics of the added mass forces, circulatory forces, and shed vortices.  
\begin{figure*}[t!]
\centering
\includegraphics[width=1\linewidth]{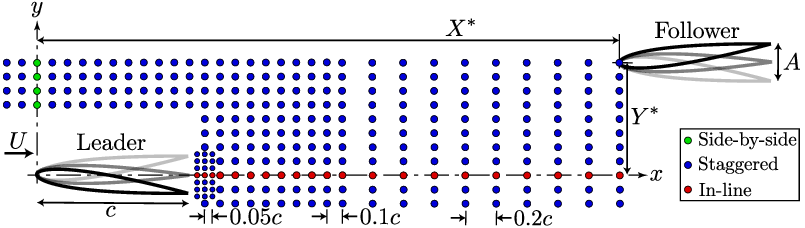}
\caption{Schematic of positions of the follower hydrofoil relative to the leader.  Shown is a grid comprised of three different spacings: a fine rectangular grid of $0.05$c spacing close to the trailing edge of the leader, ranging from $1.05 \leq X^* \leq 1.15$, $-0.15 \leq Y^* \leq 0.15$; a second region, ranging from $-0.2 \leq X^* \leq 2$, $0.5 \leq Y^* \leq 0.8$, and $1.1 \leq X^* \leq 2$, $-0.2 \leq Y^* \leq 0.4$ has a grid of $0.1$c, and a third region farther downstream ranging from $2.2 \leq X^* \leq 3.8$, $-0.1 \leq Y^* \leq 0.8$ with a more coarse grid of $0.2$c. In total there are 270 grid points.}
% Schematic of positions of the follower hydrofoil relative to the leader. Shown are two coarse rectangular grids of $0.1$c spacing ranging from $-0.2 \leq X^* \leq 2$, $0.5 \leq Y^* \leq 0.8$, and $1.1 \leq X^* \leq 2$, $-0.2 \leq Y^* \leq 0.4$, as well as a refined rectangular grid of $0.05$c spacing ranging from $1.05 \leq X^* \leq 1.15$, $-0.15 \leq Y^* \leq 0.15$.  In total there are 180 grid points.}
\label{fig:exp_setup}
\end{figure*}

Following this simple model approach, experiments were conducted on a pair of pitching hydrofoils immersed in a water channel,  forming a minimal school or minimal collective. The flow over the hydrofoils was restricted to be nominally two-dimensional. The formation of the hydrofoils was varied by manipulating the follower hydrofoil position in the streamwise and cross-stream directions, as shown in Figure \ref{fig:exp_setup}.  The dimensionless distances were normalized by the chord length as $X^*=x/c$, and $Y^*=y/c$. The leader hydrofoil was prescribed a sinusoidal pitching motion of ${\theta}_L(t) = {\theta}_0\sin(2\pi f t)$, where $f$ is the oscillation frequency, and $\theta_0$ is the pitching amplitude. The follower was pitched similarly as ${\theta}_F(t) = {\theta}_0 \sin(2\pi f t+\phi)$ with a fixed phase difference or synchrony of $\phi = \pi$ throughout the study. The peak-to-peak trailing edge amplitude of the hydrofoils can be defined as $A =2c\sin(\theta_0)$. The oscillation frequency and the dimensionless amplitude, $A^* = A/c$,  were also fixed throughout the study at $f=0.98$ Hz and $A^*=0.25$, which gives a fixed reduced frequency of $k=fc/U=1$, and a fixed Strouhal number of $St=fA/U=0.25$.  These dimensionless numbers are typical of efficient biological swimming \cite{Webb2002,Gazzola2014}. Direct force measurements were taken from each hydrofoil at every ($X^*$, $Y^*$) position, which corresponds to a total of 270 formations in the $x$-$y$ plane. For further information about the water channel setup, actuator mechanisms, sensors, the methods used, and the definition of the performance coefficients for an individual hydrofoil as well as the collective, please see \textit{Materials and Methods} and \textit{Supplementary Information}.   %%%%%%%%%%%%%%%%%%%%%%%%%%%%

%%%%%%%%%%%%%%%%%%%%%%%%%%%%

\subsection*{Follower Force Map}
In order to probe the Lighthill conjecture in two-dimensions, the \textit{relative forces} acting on the follower in the ($x$-$y$) interaction plane must be examined.  This is done by constructing a \textit{force map}, which is described below. 

\begin{figure*}[t!]
\centering
\includegraphics[width=1\linewidth]{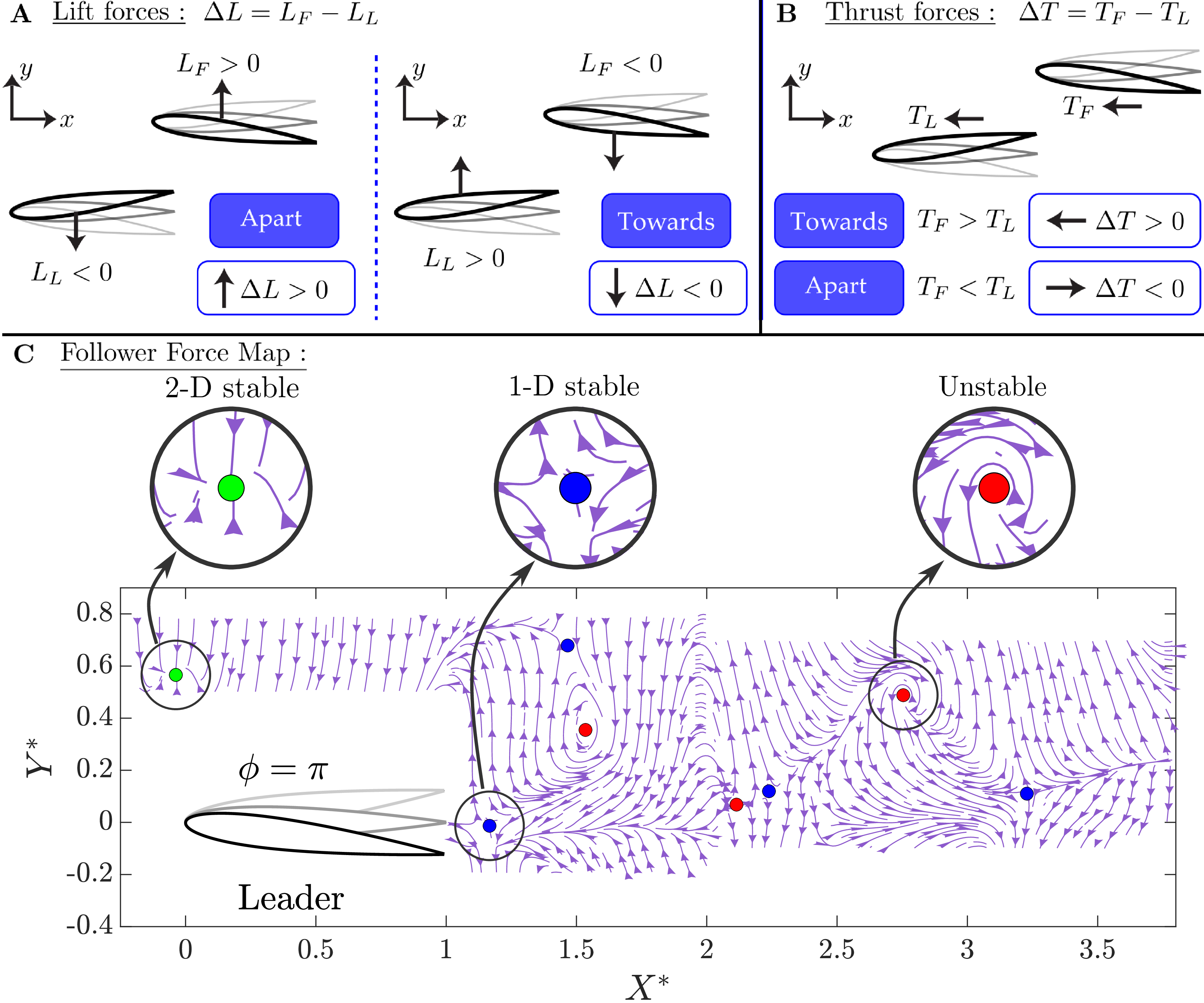}
\caption{Typical conditions leading to positive and negative relative (A) lift and (B) thrust.  (C) Follower force map with an out-of-phase synchrony between the leader and follower, i.e. $\phi = \pi$.  The arrows on the force lines indicate the direction that the follower would move in relative to the leader if it were free-swimming. The green, blue, and red circles represent the 2D stable equilibria, 1D stable/1D unstable saddle point equilibria, and the 2D unstable equilibria, respectively.}
\label{fig:force_map}
\end{figure*}

Consider a frame of reference attached to the leader as in Figure \ref{fig:exp_setup}.  The relative lift, $\Delta L$, in the cross-stream direction and the relative thrust, $\Delta T$, in the streamwise direction are defined simply as a difference between the forces acting on the two hydrofoils as $\Delta T = T_F - T_L$ and $\Delta L = L_F - L_L$, where forces acting on the leader and follower hydrofoils are denoted with $(.)_L$ and $(.)_F$, respectively.  Figure \ref{fig:force_map}A and \ref{fig:force_map}B show the relative force conditions that lead to the follower either moving towards or moving away from the leader in the streamwise ($x$) and cross-stream ($y$) directions. 

% \begin{figure*}[t!]
% \centering
% \includegraphics[width=1\linewidth]{figures/ForceMap_Resub.eps}
% \caption{New force map.}
% \label{fig:new_force_map}
% \end{figure*}

First, consider the relative lift for the positive $x$-$y$ plane. The follower is pushed away from the leader in the cross-stream direction ($\updownarrow$) when the relative lift force is greater than zero (Figure \ref{fig:force_map}A).  This condition arises either when lift forces acting on the foils are in the same direction and $L_F > L_L$, or when they are acting in opposite directions and pointing away from each other ($L_L<0 \downarrow$, $L_F>0 \uparrow$). In contrast, the follower is pulled towards the leader when the lift forces are acting in the same direction and $L_L > L_F$ or acting in opposite directions and pointing towards each other ($L_L>0 \uparrow$, $L_F<0 \downarrow$).

Next, consider the relative thrust force in the positive $x$-$y$ plane.  A positive relative thrust force ($\Delta T>0$) acts to move the follower towards the leader, which arises when $T_F>T_L$. In contrast, when $T_L>T_F$ the relative thrust force is negative ($\Delta T < 0$), and the follower moves away from the leader in the streamwise direction as shown in Figure \ref{fig:force_map}B.  If $T_L = T_F$, then the leader and follower swim at the same speed and do not move closer or apart. 

To visualize the directions of the relative forces acting on the follower throughout the $x$-$y$ plane we constructed a force map, which is a novel visualization made up of force lines (Figure \ref{fig:force_map}C).  Put simply, the force map conveys \textit{the direction that the follower would move in as observed by the leader.}  The force map is constructed with the origin located at the leading edge of the leader and a relative force vector ($\mathbf{F}_\text{rel} = -\Delta T \mathbf{\hat{x}} +  \Delta L \mathbf{\hat{y}}$) is determined at each of the measurement positions detailed in Figure \ref{fig:exp_setup}, such that a relative force vector field is created.  Force lines are then graphed as lines that are everywhere tangent to the local relative force vector field, analogous to streamlines. This novel visualization tool developed here, \textit{for the first time}, uses time-averaged force data to visualize the stability characteristics of constrained foils, yet it will be shown to be indicative of the stability characteristics of \textit{unconstrained, freely-swimming foils}. 

%%%%%%%%%%%%%%%%%%%%%%%%%%%%

%%%%%%%%%%%%%%%%%%%%%%%%%%%%
\subsection*{Observed Equilibria}
The force map reveals eight critical points or equilibria where the relative force vector is equal to zero, that is,  $\Delta T=0$ and $\Delta L=0$, which are marked by green, blue, and red circles in Figure \ref{fig:force_map}C. The first equilibrium point is located at $(X^*, Y^*)=(0, 0.6)$ where the leader and follower are interacting in a \textit{side-by-side} formation. Interestingly, as the force lines merge at this point, their direction indicates that this equilibrium point is a \textit{stable sink point} (green circle) in two-dimensions. Therefore, when any perturbations move the follower away from this point, forces will arise to return the foil back to this location. Previous low Reynolds number simulations have also shown that side-by-side formations of bending foils are two-dimensionally stable \cite{dai2018stable}, at least for foils constrained in the cross-stream direction. Another critical point is located at $(X^*,\, Y^*)=(1.2,\, 0)$ in the leader's wake where the follower is directly \textit{in-line} with the leader. This represents an equilibrium point that is stable to streamwise perturbations, but unstable to cross-stream perturbations, that is, an \textit{unstable saddle point} (blue circles). In fact, there are repeating unstable saddle points in the wake zone ($-0.2 \leq Y^* \leq 0.2$) that are spaced one wake wavelength apart located at $X^*= 2.2$ and $3.2$. Previous studies \cite{becker2015hydrodynamic,Ramananarivo2016,newbolt2019flow} have shown that these equilibria are one-dimensionally stable in the streamwise direction, however, we now reveal that they are unstable in the cross-stream direction. Beyond unstable saddle points there are also equilibrium points that are \textit{unstable source points} (red circles). These are locations of unstable equilibria in both the streamwise and cross-stream direction with force-lines radiating outward from them. 

%%%%%%%%%%%%%%%%%%%%%%%%%%%%

%%%%%%%%%%%%%%%%%%%%%%%%%%%%
\subsection*{Flow Mechanisms Behind the Stable Formation}
\begin{figure*}[t!]
\centering
\includegraphics[width=1\linewidth]{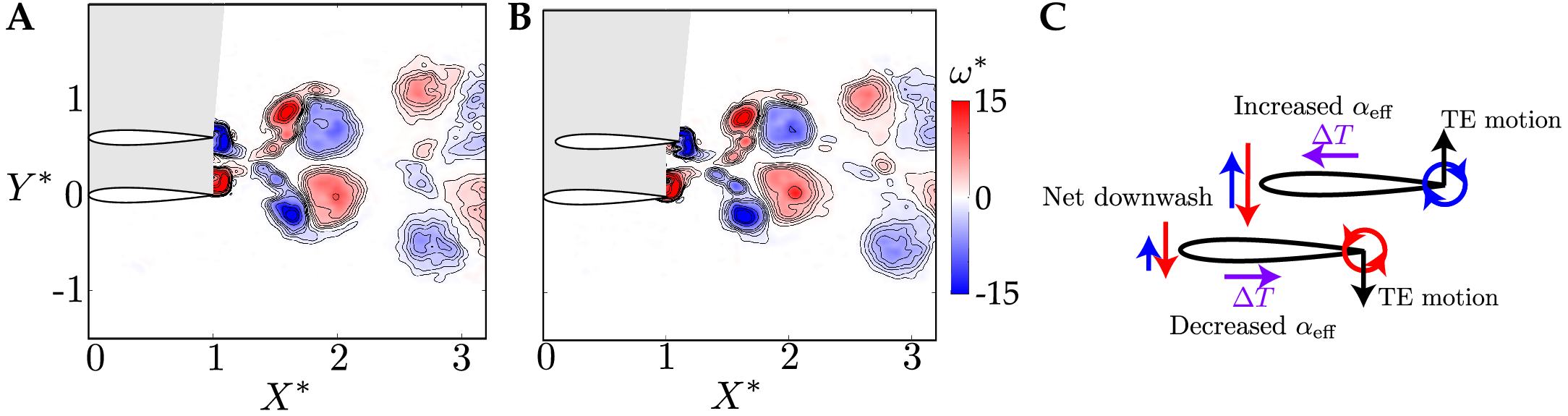}
\caption{(A) Vorticity field for the stable side-by-side formation ($X^* = 0$, $Y^* = 0.6$) at $t^* = 0.5$ when the foils are pitching away from each other. (B) Vorticity field for the formation of $X^* = 0.1$, $Y^* = 0.6$ at $t^* = 0.5$. (C) Schematic of proposed trailing-edge vortex mechanism responsible for a restorative force back to the stable side-by-side formation.}
\label{fig:sidebysidewakes}
\end{figure*}
To understand the flow mechanisms behind the stable side-by-side formation particle image velocimetry (PIV) measurements are employed.

Figure \ref{fig:sidebysidewakes}A presents the vorticity shed from the leader and follower at the stable side-by-side equilibrium at a dimensionless time of $t^* = t/T = 0.5$ where the foils are pitching away from each other.  The forming clockwise vortex of the follower and counterclockwise vortex of the leader mutually induce each other in a way that slows their downstream advection.  These vortices then pair with their counter-rotating counterparts shed a half cycle later. The pairs mutually induce away from the symmetry line between the foils, leading to momentum jets behind the foils that deflect away from the symmetry line.  The deflected jet and vortex pairing mechanism were first observed in \cite{quinn2014unsteady} and later in subsequent work \cite{Bao2017}. The cross-stream stability of the side-by-side formation is mediated by a balance of the wake-induced forces from the deflected jet and the quasi-steady forces, that is, the body-body flow interaction between the foils, which has been established in \cite{kurt2019swimming,han2023revealing}.  

What has not been established is the mechanism that generates \textit{restorative streamwise forces} for the leader and follower to bring them back into equilibrium when one is perturbed downstream of the other. For instance, when the follower is located slightly downstream of the side-by-side formation (Figure \ref{fig:sidebysidewakes}B) there are the same vortex pairs as observed at the equilibrium formation only with a slight asymmetry, which brings the forming counterclockwise vortex of the leader closer to the leading edge of the follower (Figure \ref{fig:sidebysidewakes}C). This leads to an increase in the effective angle of attack and thrust of the follower.  Concurrently, the forming clockwise vortex of the follower is farther from the leading edge of the leader than at the equilibrium formation.  This decreases the effective angle of attack and thrust of the leader.  Taken together, the increased thrust of the follower and decreased thrust of the leader act to restore them back to their equilibrium formation. 

%%%%%%%%%%%%%%%%%%%%%%%%%%%%

%%%%%%%%%%%%%%%%%%%%%%%%%%%%
\subsection*{Fluctuating Forces} \label{sec:fluctuations}
Beyond the mean forces that define the force map for the $x$-$y$ interaction plane, the thrust and lift forces also exhibit fluctuations throughout a foil's pitching cycle. In order to understand this aspect, we analyzed these fluctuations for both the leader and follower as presented in Figure S3 in the \textit{Supplementary Information}. The lift fluctuations are found to be an order of magnitude higher than the thrust fluctuations for the constrained foils, as previously observed in \cite{li2019energetics} for a pair of interacting fish through 3D computations. At the stable side-by-side formation, the force fluctuations increase by 30-40\% over an isolated hydrofoil. Around the in-line unstable saddle point at $(X^*,\, Y^*)=(1.2,\, 0)$, the force fluctuations on both foils are, surprisingly, reduced compared to an isolated hydrofoil.  At the other equilibria, the fluctuations stay effectively at about the same levels as an isolated hydrofoil.  Also, in the wake of the leader around $1.5\leq X^* \leq 2$, there is an increase in the follower's lift and thrust fluctuations by 20-40\%.
%%%%%%%%%%%%%%%%%%%%%%%%%%%%

%%%%%%%%%%%%%%%%%%%%%%%%%%%%
\subsection*{Dynamic Schooling Interactions}
Through the use of a novel force map, only one \textit{two-dimensionally stable} equilibrium point for a side-by-side formation has been discovered within the interaction plane. However, the force map assumes two simplifications compared to a freely-swimming, dynamical system: (1) it filters out the time-varying forces by using time-averaged data, and (2) it removes the dynamic recoil motion of freely swimming foils due to the fluctuating forces described in \textit{Fluctuating Forces}.  Dynamic recoil motion occurs when a two-dimensionally unconstrained pitching hydrofoil, for example, pitches through its downstroke.  During this stroke a positive lift force is generated, leading to a \textit{recoil} motion where the body will heave upward in response to the lift.  In this sense, the recoil motion introduces heaving that lags the pitching motion by nearly 180$^o$, which is known to lower the thrust production \cite{Buren2019} and may alter the lift and, \textit{importantly}, the stability of equilibria. In order to explore the effect of these dynamic recoil motions on the stability of the side-by-side equilibrium, to further verify the findings from the force map, and to determine the free-swimming performance benefits of the side-by-side formation, freely swimming experiments are developed, and companion simulations were performed.

% \sout{However, this result neglects the effect of dynamic recoil motion of freely swimming bodies due to the fluctuating forces described before.} 

Experimental measurements of the two-dimensional stability of freely swimming foils is achieved, \textit{for the first time}, by mounting each foil actuation mechanism on a novel double air bearing stage for nearly friction-less motion, and with on-board batteries and wireless communication to eliminate forces due to electronic wiring. Exceptional care is taken to align, level, and counter-bend the air-bearing rails in order to minimize these sources of non-hydrodynamic forces acting on the foils. For more details on the unconstrained experimental approach, please see \textit{Unconstrained Foil Experiments}. To verify that these novel experiments are measuring the actual hydrodynamic forces acting on interacting foils instead of being corrupted by non-hydrodynamic forces, such as settling to a false equilibrium due to rail bending, we have reproduced results of the streamwise stable equilibria discovered in \cite{newbolt2019flow}. These validation results can be found in the \textit{Supplementary Information}. Furthermore, in a second experiment, also found in the \textit{Supplementary Information}, the follower starts at a range of in-line formations, but it is unconstrained in both the streamwise and cross-stream directions.  This experiment examines the \textit{two-dimensional} stability of in-line formations, which are predicted by the force map to be unstable saddle points. Indeed, the unconstrained measurements verify that the follower diverges from the in-line formations, as predicted, and never settles into a two-dimensionally stable formation downstream of the leader.

\begin{figure*}[h]
\centering
\includegraphics[width=1\linewidth]{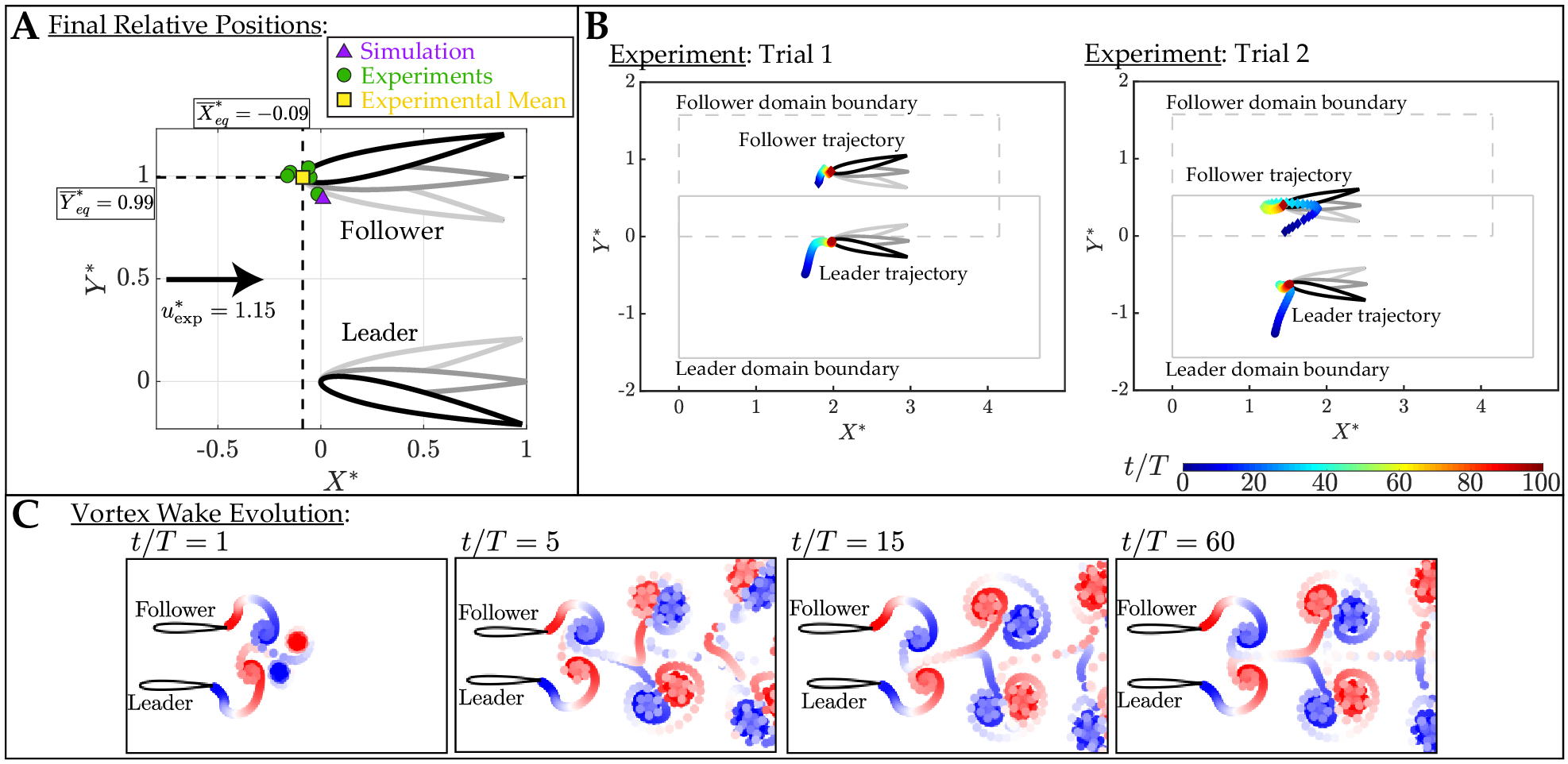}
\caption{\textbf{A}: Equilibrium positions for two fully unconstrained hydrofoils, free to move in the $X^{*}$-$Y^{*}$ plane with a phase synchrony of $\phi = 180^{\circ}$ graphed in a relative frame of reference. The equilibrium positions from individual experiments are represented by the green circles while the experimental mean equilibrium position  $(\overline{X}^{*}_{eq},\overline{Y}^{*}_{eq}) = (-0.09,0.99)$ is represented by the yellow square, and the numerical equilibrium position $(X^*_{eq},\, Y^*_{eq}) = (0,\, 0.90)$ is represented by the purple triangle. The resulting stable side-by-side formation achieves a normalized swimming speed in the experiments of $u_\text{exp}^{*} = 1.15$, which is 15\% higher than that of an isolated swimmer.
\textbf{B}: Trajectories for two experimental trials graphed in an absolute frame of reference. The markers are colored from blue to red based on the dimensionless time, and each marker represents the cycle-averaged positions of the swimmers. After around 70 cycles, the relative position of the swimmers remains constant; hence an equilibrium configuration is achieved. The two trials show that equilibrium formations can be achieved throughout the experimental domain.    
\textbf{C}: Simulated vortex wake evolution for two free-swimming hydrofoils mirrors the wake measurements of constrained foils seen in Figure \ref{fig:sidebysidewakes}.}
\label{fig:sidebyside_freeswimming}
\end{figure*}

Next, we present experimental measurements and companion simulations of two pitching hydrofoils that start near the stable equilibrium point and are free to move in the $x$-$y$ plane. Companion simulations are run with a flow solver calculating the unsteady potential flow, i.e., the inviscid, irrotational, and incompressible flow produced by a pair of pitching hydrofoils.  Further details of the numerical method and parameters used can be found in the \textit{Materials and Methods} section. In both the simulations and experiments, multiple trials of varying initial conditions consistently evolve to approximately the same final states (Figure  \ref{fig:sidebyside_freeswimming}A) with a mean experimental equilibrium position of $(\overline{X}^*_{eq},\, \overline{Y}^*_{eq}) = (-0.09,\, 0.99)$ and a numerical equilibrium position of $(X^*_{eq},\, Y^*_{eq}) = (0,\, 0.90)$, within 10\% of the separation distance measured in the experiments.  While the final state of the experimental foils has a consistent relative spacing from trial-to-trial, two example time-varying trajectories (Figure \ref{fig:sidebyside_freeswimming}B) show the time evolution of the foils' absolute positions within the experimental domain can be quite different.  This supports the assertion that the experiments are not contaminated with false equilibrium positions that are artifacts of rail bending, tilting, etc. Both the experiments and simulations show that a side-by-side stable formation does indeed exist even when there is dynamic recoil motion and for both two-dimensional flows (simulations) and three-dimensional flows (experiments; $AR = 3$). Since the simple potential flow simulations can closely predict the equilibrium formation, this further shows that the physics driving the stable side-by-side formation is dominated by inviscid mechanisms. This supports the proposed inviscid mechanism discussed in \textit{Flow Mechanisms Behind the Stable Formation}. Figure \ref{fig:sidebyside_freeswimming}C shows the wake flows generated by the interacting freely-swimming pitching foils.  These simulated wake flows are characteristically the same as those highlighted in the PIV experiments presented earlier.  

To better understand the connection between the kinematics and the stable equilibrium position, five simulation cases with varying dimensionless amplitude are considered (Table \ref{TAB:dynamicschool}), where Case II is the simulation data plotted in Figure \ref{fig:sidebyside_freeswimming}A. For each case the freely-swimming foils are found to converge to an equilibrium point in a side-by-side formation. As the amplitude is increased the Strouhal number remains nearly constant at $St \approx 0.3$ since the swimming speed is an output of freely-swimming simulations and $U$ scales with the amplitude and the pitching frequency as $U\propto fA$ \cite{bainbridge1958speed, saadat2017rules, moored2019inviscid}. Since the speed increases with increasing amplitude, the reduced frequency decreases over the range $0.9 \geq k \geq 0.29$. The data show that in each equilibrium formation, the cross-stream foil spacing increases as the reduced frequency decreases. This is the same trend observed previously for a pitching foil in ground effect \cite{kurt2019swimming}. 
\begin{table}%[tbhp]
\centering
\caption{Simulation input and output data for five cases of varying amplitude. The equilibrium position is denoted as $(X^*_\text{eq}$, $Y^*_\text{eq})$ and the origin is defined at the leading edge of the leader. Data for an isolated swimmer are denoted with a superscript $(\cdot)^\text{iso}$.}
\begin{tabular}{cccccccccccc}%{lrrrrrrrr}
Parameters & Case I &  Case II & Case III & Case IV & Case V\\
\midrule
$A^*$   & 0.3    &  0.33  & 0.4    & 0.45  & 0.5 \\
$St$    & 0.27 & 0.3 & 0.28 & 0.32 & 0.29 \\
$k$     & 0.9 & 0.87 & 0.70 & 0.64 & 0.29 \\
$(X^*_\text{eq}$, $Y^*_\text{eq})$  & (0, 0.68) &  (0, 0.9) & (0, 0.99) & (0, 1.14)  & (0, 1.28) \\
$u^*$   & 1.10 & 1.08 & 1.07 & 1.06 & 1.05 \\
% $C^*_{T,a}$ & 1.1834& 1.1624 & 1.1429 & 1.0985 \\
% $C^*_{P,a}$ & 1.1120& 1.1010 & 1.0853 & 1.0660 \\
$\eta^*$& 1.09 & 1.06& 1.06 & 1.05 & 1.04 \\
% $C^{\text{iso}}_{T,a}$ &  1.4937  & 1.4778  & 1.4622 & 1.4441\\
% $C^{\text{iso}}_{P,a}$ & 4.5204 & 4.4757 &  4.4566  & 4.660\\
$\eta^{\text{iso}}$ & 0.32 & 0.31 & 0.32 &  0.31  & 0.3\\
$u^{\text{iso}}$ [m/s] & 0.097& 0.102    &  0.128   &   0.142    & 0.157\\
% $COT^*$ & 5.0261& 4.9198 & 4.8421 &        \\
\bottomrule
\end{tabular}
\label{TAB:dynamicschool}
\end{table}

%%%%%%%%%%%%%%%%%%%%%%%%%%%%

%%%%%%%%%%%%%%%%%%%%%%%%%%%%
\subsection*{Schooling Performance}
Beyond probing the Lighthill conjecture in two dimensions, hydrofoil performance was also measured for the free-swimming experiments and simulations, as well as for the constrained experiments for formations throughout the interaction plane. First, the free-swimming experiments show that swimming side-by-side increases the swimming speed of the pair of foils by 15\% compared to an isolated foil.  The simulations also show a swimming speed benefit, though reduced from the experiments, of $5-10\%$ for a pair of foils with more compact formations leading to increased gains. The reduced speed benefit in the simulations is likely due to their modelling simplifications as two-dimensional, inviscid flow. Still, the simulations capture the same trends as the experiments and provide further information showing that the efficiency can also increase by $4-9\%$ for a freely-swimming pair of foils. 
% These performance increases originate from increases in the added mass of the foils as established recently \cite{Mivehchi2021}.  

The freely-swimming performance data from the experiments and simulations, while important, only can be measured at equilibrium formations.  However, the constrained foil measurements used to generate the force map can be leveraged to investigate the performance landscape at all possible formations over the interaction plane at \textit{out-of-equilibrium} conditions. This can provide a map that can be used to highlight high-performance zones and their connection to stable formations.

\begin{figure*}[t!]
\centering
\includegraphics[width=1\linewidth]{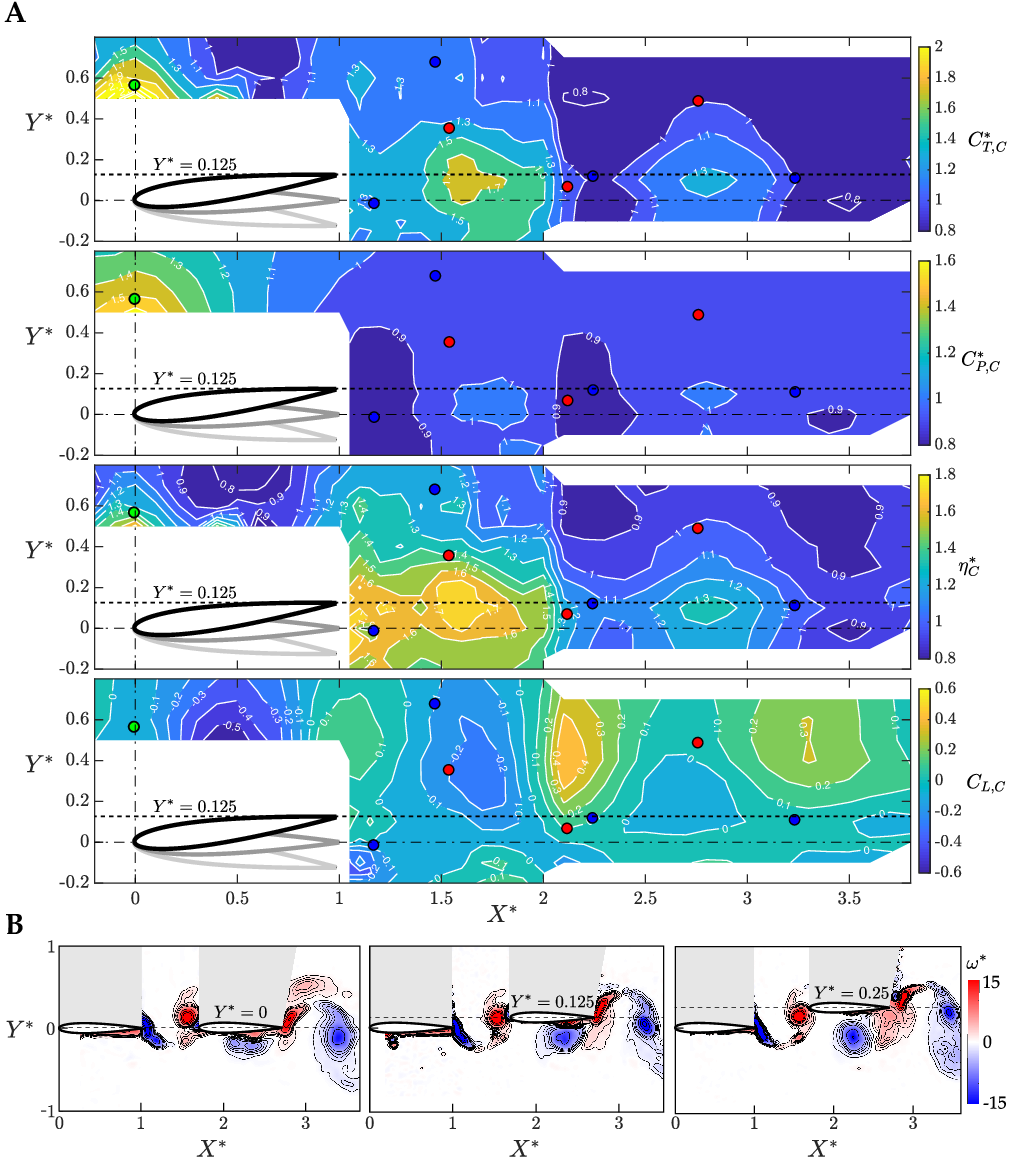}
\caption{\textbf{A}: Contour maps arranged from top to bottom of normalized collective thrust, power, efficiency, and collective lift. The black dashed lines show the locations corresponding to $Y^*=0.125$, which is the maximum trailing edge position of the leader foil.  The leader position, size, and amplitude of motion are shown for reference. The green, blue, and red circles represent the 2D stable, 1D stable/1D unstable, and unstable equilibria, respectively, from Figure \ref{fig:force_map}.
\textbf{B}: Vorticity fields at the end of a cycle during the leader's up-stroke and the follower's down-stroke for, from left to right, the in-line formation of $X^*=1.7$ and $Y^*=0.0$, the staggered formation of $X^*=1.7$ and $Y^*=0.125$, and the staggered formation of $X^*=1.7$ and $Y^*=0.25$.}
\label{fig:collective_performance}
\end{figure*}

Figure \ref{fig:collective_performance}A presents the normalized collective thrust, $C^*_{T,C}$,  power, $C^*_{P,C}$, efficiency, $\eta_C^*$, and the collective lift, $C_{L,C}$, as a contour map of $(X^*$, $ Y^*)$. The normalized performance metrics compare the collective performance of the leader-follower pair with that of two \textit{isolated} hydrofoils. Note that in contrast to the relative forces that were discussed in the previous section, a \textit{collective} lift of zero does not necessarily mean zero lift for the foils individually.

Generally, the normalized collective thrust stays within the range of $1 \leq C_{T,C}^* \leq 2$, except for sparse formations where the follower is greater than one chord downstream of the leader. This indicates that the collective performs better than two hydrofoils in isolation throughout much of the interaction plane considered here. Previous work has also observed similar thrust enhancements, albeit for a limited set of data of one side-by-side and two staggered formations at $Y^* = 1$ \cite{Huera-Huarte2018}. The peaks in collective thrust can be grouped into two regions. First, there is a region enclosing the side-by-side formations along the line of $X^*=0$ (vertical dash-dot line), where the collective is found to achieve $40-100\%$ higher thrust than in isolation. Recent work has established \cite{Mivehchi2021} that the thrust increase for out-of-phase pitching foils in this region originates from an increase in their added mass.  It was determined that the added mass thrust dominates the thrust production of pitching foils, and thereby wake effects play no significant role in their thrust increase. However, for heaving or combined heaving and pitching foils, wake effects and circulatory forces, in general, are expected to be important for thrust production in a side-by-side formation \cite{Ayancik2020}. A second region corresponds to the direct wake interactions enclosing the in-line and slightly-staggered formations within the region of $0\leq Y^*\leq 0.3$. For the in-line formation at $X^* = 1.6$, the collective thrust reaches a $77\%$ peak increase over the hydrofoils in isolation, whereas the thrust reaches a $87\%$ peak increase for the slightly-staggered formation along the $Y^*=0.125$ line.  This line is where direct wake vortex impingement onto the follower occurs, as shown in Figure \ref{fig:collective_performance}B. It has already been established that wake-body impingement interactions generate increased thrust through an increase in the effective angle of attack of the follower and are therefore circulatory in nature \cite{Boschitsch2014,muscutt2017performance}. One wake wavelength downstream of the primary thrust peak at $X^* = 2.8$ there is a second thrust peak reaching a reduced collective thrust benefit from the primary peak of a 39\% increase over two isolated foils. This highlights that the highest performance benefits of schooling occur in compact formations. For near wake interactions where the collective is in in-line or staggered formations ($X^* > 1.1$ and any $Y^*$), the normalized collective power exhibits little variation from the isolated case, whereas the side-by-side formations result in up to a 60\% increase in power. 

The collective efficiency is observed to increase by $10-40\%$ over that of isolated foils for the side-by-side interaction region, which has been observed previously \cite{Huera-Huarte2018}. In the in-line interaction region at $X^* = 1.6$, even higher peak efficiency increases are identified with up to a $73\%$ increase for in-line interactions and an $84\%$ increase for slightly-staggered formations along the $Y^*=0.125$ dashed line. It is clear that while both the side-by-side and in-line interaction regions see comparable thrust increases, the efficiency increase in the side-by-side interaction is tempered by a concurrent rise in power, whereas the efficiency increase in the in-line interaction region is solely driven by the increase in thrust. Some previous work has not observed an efficiency benefit for staggered formations with $Y^* = 1$ \cite{Huera-Huarte2018}, which is reflected in the small efficiency gains around 10-20\% on the edge of the current experimental domain at $Y^* = 0.8$.  This efficiency benefit would presumably decay further towards 0\% at $Y^* = 1$. However, in the near-wake interactions probed in the current study with $Y^* \leq 0.8$ it is revealed that the greatest efficiency benefits are observed for slightly staggered formations. Moreover, as seen in the thrust data, there is a secondary efficiency peak located one wake wavelength downstream of the primary efficiency peak, but with a reduced maximum efficiency increase of 37\%.

For side-by-side formations, where a stable equilibrium point is located, the collective lift generation is found to be negligible. This means that when two individuals are swimming in this stable formation, the formation is, in fact, \textit{super-stable}, that is, the relative distances between the swimmers do not change \textit{and} the pair of swimmers will remain swimming forward without a collective drift to one side or another.  Likewise, around the in-line interaction region, where unstable saddle points are located, collective lift generation is found to be negligible as well.

% \kwm{The thrust and efficiency benefits are observed to be tempered due to the dynamic recoil motions of both foils.}

%%%%%%%%%%%%%%%%%%%%%%%%%%%%

%%%%%%%%%%%%%%%%%%%%%%%%%%%%
\section*{Discussion and Conclusions}
For the first time, we have discovered that a side-by-side formation of pitching foils is not only \textit{two-dimensionally} stable, but also two-dimensionally \textit{super-stable} where their relative distance stays constant \textit{and} the school as a whole does not have forces acting on it to drift to one side or the other during locomotion.  Indeed, the school is shown to be super-stable for freely-swimming foils, which also enjoy modest speed and efficiency gains over swimming in isolation. This provides new evidence that the Lighthill conjecture \cite{lighthill1975mathematical} may play a role in school formation and spatial patterns of swimmers, even if only in a statistical sense akin to birds in a V-formation \cite{portugal2014upwash}.  Indeed, it has been observed \cite{ashraf2017simple} that above a critical swimming speed, tetra fish organize in a side-by-side lineup (or phalanx formation as described in the study) and that they enjoy some energetic benefit.  Our findings support this previous work and provide a new viewpoint that perhaps the self-organization of the tetra fish is passive in nature and not an active control strategy by the fish.  

While this is a provocative result, there are many differences between schooling pitching foils and schooling tetra fish that make it difficult to draw a direct conclusion about the fish.  One important difference is that the fish are composed of a body and fins, with their caudal fin undergoing a combined heaving and pitching motion.  However, recent work has shown that both purely heaving and purely pitching foils experience one-dimensional stability in an in-line formation \cite{Heydari2020}, so perhaps the difference in kinematics is not consequential for the existence of stable equilibria.  Still, further work should aim to examine the passive stability of truly fish-like swimmers to answer this question directly.  

Another complicating factor is that it has been argued that three-dimensional swimmers in an infinite school experience breakdown of their shed wake vortices \cite{Daghooghi2015}, thereby disrupting the coherent vortex-body interactions that drive the in-line and staggered formation interactions.  The side-by-side interactions, though, are dominated by oscillating dipole flow fields produced from motions of the nearby bodies, which cannot break down like wake vortices.  This may mean that a stable side-by-side formation and its performance benefits would hold even for three-dimensional swimmers and/or dense schools.  Further work in this direction is also warranted to provide some understanding of how the two-body two-dimensional interactions of the current study translate to many-bodied and three-dimensional interactions.  

Previous work \cite{becker2015hydrodynamic,Ramananarivo2016,newbolt2019flow} have shown that in-line formations have multiple \textit{one-dimensionally} stable equilibria, while the current results show that those equilibria are, in fact, unstable in the cross-stream direction. However, this does not indicate that these one-dimensionally stable points are irrelevant. Instead, this work highlights that the degree of stability falls along a spectrum.  For instance, a fish swimming in the wake of another fish may only need to actively control its cross-stream position in order to maintain a schooling formation, which requires less control effort than actively controlling two degrees of freedom, but more effort than controlling none.  While higher degrees of stability can relieve the need for control strategies for swimmers and lead to completely passive self-organization, lesser degrees of stability may more subtly sculpt the schooling patterns observed in fish by influencing the trajectory manifolds or the statistical positioning of swimmers.  Beyond the translational stability of formations, it is unclear whether the orientation of a swimmer will also be stable to perturbations since it is unstable, at least for some synchronizations and formations \cite{Gazzola2011}. 

Interestingly, when the thrust and efficiency performance are considered, the ideal formation for maximizing performance is not the super-stable side-by-side formation, though modest thrust and efficiency benefits can be reaped in this formation.  The optimal thrust and efficiency performance occurs for a slightly staggered formation, which gives rise to interesting questions about whether animals swim in energetically optimal formations through more attentive control or in more stable formations with fewer performance benefits. 

This study provides a rich understanding of the interplay of stability and performance in schooling pitching foils. These findings reveal hypotheses for understanding biological schooling and also provide insights that aid in the design of multi-finned or schools of bio-inspired machines.

%%%%%%%%%%%%%%%%%%%%%%%%%%%%

%%%%%%%%%%%%%%%%%%%%%%%%%%%%

\matmethods{\subsection*{Constrained Foil Experiments}
Experiments were conducted for a minimal collective consisting of a pair of hydrofoils constrained in space and immersed in a closed-loop water channel, oscillating under prescribed sinusoidal pitching motions about an axis located $8.4$ mm behind their leading edge.  A constant flow speed of $U=0.093$ m/s was imposed, which gives a chord-length based Reynolds number of $Re=9,$950. The flow over the hydrofoils was restricted to be nominally two-dimensional by the placement of a surface and splitter plate near the hydrofoil tips. Two identical hydrofoils with a rectangular planform and NACA 0012 cross-section were designated as the leader and follower. Each hydrofoil had a chord length of $c=0.095$ m and a span length of $s=0.19$ m, which gives an aspect ratio of $AR=2$.
\\
Particle Image Velocimetry measurements of the flow field were performed in the horizontal ($X^*$-$Y^*$) plane at the mid-span of the foils. Phase-averaged results were calculated from a total of 100 measurements for each flow-field. A total of 16 distinct phases ($0, \frac{\pi}{8}, \frac{\pi}{4}, ... \frac{15\pi}{8}$) were captured for all schooling formations. The camera used is an Imager sCMOS (2560$\times$2560 pixels) paired with lens of 50mm of focal length and f-number $f_\# = 2.8$. A magnification factor of 0.135 yields a field of view of (3.64c $\times$ 3.07c). Seeding particles are hollow metallic coated plastic spheres of 11 $\mu$m in diameter were illuminated by a 200 mJ/pulse Nd:YAG laser. Multi-pass, cross-correlation processing of the raw images was employed to obtain the resulting vector fields, with a final interrogation window of 48$\times$48 pixels.

%%%%%%%%%%%%%%%%%%%%%%%%%%%%

%%%%%%%%%%%%%%%%%%%%%%%%%%%%

\subsection*{Force Measurements}
An ATI Nano43 six-axis force sensor was used to measure the thrust, lift and pitching moment acting on each hydrofoil. An optical encoder recorded the angular position, which was then used to compute the angular velocity, $\dot{\theta}$, for each hydrofoil. The total instantaneous power input was then calculated as $P_T(t)=M_{\theta}\dot{\theta}$, where $M_{\theta}$ denotes the pitching moment. Here, the inertial power was determined from the same experiments conducted in air, and was subtracted from the total power, $P_T(t)$, to calculate the instantaneous power input to the fluid, $P(t)$. Force measurements were taken for 100 oscillation cycles from the leader and follower, and each experiment was repeated 10 times. The time-averaged values were calculated for each of these trials, and their mean from 10 trials was calculated to determine the time-averaged total thrust, lift, and power. The profile drag was measured for the static foil in an imposed flow over 20-second intervals. Net thrust was determined by subtracting the profile drag acting on the hydrofoils from the time-averaged thrust as follows, $\overline T_{\text{net}} = \overline T - \overline D$. The definitions of the coefficients of net thrust, $C_T$, lift, $C_L$, and power, $C_P$, and efficiency, $\eta$, are given as follows for the individual hydrofoils:
\begin{equation} \label{eq:coefficients}
\begin{aligned}
C_T = \frac{\overline T_{\text{net}}}{\frac{1}{2}\rho {U}^2 c s}, \quad  C_L = \frac{\overline L}{\frac{1}{2}\rho {U}^2 c s},\\
C_P = \frac{\overline P}{\frac{1}{2}\rho {U}^3 c s},  \quad  \eta = \frac{C_T}{C_P},
\end{aligned}
\end{equation}
where $\rho$ is the fluid density and $s$ is the span-length of the hydrofoils.
Here, we also report collective performance parameters, that is, the average performance from the leader and the follower. The collective force and power coefficients, as well as the collective efficiency, are denoted with a $C$ subscript, and they are defined as,
\begin{equation}
\begin{aligned}
C_{T,C} = \frac{\overline{T}_L + \overline{T}_F}{\rho {U_\infty}^2 c s}, \quad C_{L,C} = \frac{\overline{L}_L + \overline{L}_F}{\rho {U_\infty}^2 c s},\\   C_{P,C} = \frac{\overline{P}_L + \overline{P}_F}{\rho {U_\infty}^3 c s}, \quad  \eta_C = \frac{C_{T,C}}{C_{P,C}}. 
\end{aligned}
\end{equation}
Note that, here, the performance coefficients were defined with combined propulsor area, that is $2cs$, cancelling the one-half in the denominator. Collective thrust and power coefficients, and efficiency were reported as normalized values with the corresponding isolated hydrofoil performance metric for comparison, and defined as follows,  
\begin{equation}
C_{T}^* = \frac{C_{T,C}}{C^\text{iso}_{T}},  \quad  C_{P}^* = \frac{C_{P,C}}{C^\text{iso}_{P}}, \quad \eta^* = \frac{\eta_C}{\eta^\text{iso}}. 
\end{equation}
Here, the collective performance metrics are compared with the collective of \textit{two} isolated hydrofoils, $C^\text{iso}_{T,C} = C^\text{iso}_{T}$. The isolated net thrust, drag, power and efficiency are $C^\text{iso}_{T} = 0.10 \pm 0.015$, $C^\text{iso}_{D} = 0.03 \pm 0.002$, $C^\text{iso}_{P} = 0.66 \pm 0.0008$, and $\eta^\text{iso} = 0.15 \pm 0.022$, respectively.

%%%%%%%%%%%%%%%%%%%%%%%%%%%%
%%%%%%%%%%%%%%%%%%%%%%%%%%%%

\subsection*{Unconstrained Foil Experiments}\label{sec:free_swmimming_exp}
The unconstrained foil experiments were conducted to simulate a minimal school of two hydrofoil swimmers where both are unconstrained (free-to-move) in the $X^*$-$Y^*$ plane independent from each other. Similar to the constrained foil experiments, two identical hydrofoils with a rectangular planform, NACA0012 cross-section, and aspect ratio $AR=3$ were designated as the leader and follower. 

In free swimming, the swimming speed and its derivative dimensionless numbers ($Re$, $St$ and $k$) are \textit{dependent variables}, which depend upon the balance of thrust and drag of the swimmer. The thrust-drag balance is represented by the  Lighthill number \cite{moored2019inviscid}, $Li = C_D S_{wp}$, where $C_D$ is the drag coefficient of the swimmer and $S_{wp}$ is a ratio of the drag-generating wetted area to the propulsive planform area of the swimmer. For the hydrofoils in this experiment $S_{wp} \approx 2$ since the whole hydrofoil is both the drag and propulsive source.  Since this is the case, if the same prescribed kinematics from the constrained-foil measurements are used, then the free-swimming $St$ and $k$ will be much lower than the constrained experiments.  In order to achieve $St$ and $k$ comparable to the constrained-foil experiments, drag generating bodies are attached to the streamwise carriage. Three NACA 0012 foils of $AR = 3$ at zero angle of attack are used for this purpose (see \textit{Supplementary Material}). The drag-producing bodies were positioned sufficiently far from the leader and the follower to prevent them from disturbing the flow around the foils, and constrained in the cross-stream direction to ensure that they are generating viscous drag (skin-friction) only that is virtually independent of the system dynamics. If the drag-producing bodies were free to move in the cross-stream direction, their lateral motion would generate time-dependent thrust and form-drag forces. 

Similar to the constrained foil experiments, the foils were prescribed with sinusoidal pitching motions by using a servo motor. The foils receive start/stop commands from a computer via an infrared signal. Each servo motor is controlled by an Arduino MEGA2560. A detailed description of the equipment and data processing can be found in the \textit{Supplementary Material}.

%%%%%%%%%%%%%%%%%%%%%%%%%%%%
%%%%%%%%%%%%%%%%%%%%%%%%%%%%

\subsection*{Numerical Methods}\label{sec:num_methods}
To model the flow over a foil unconstrained both in the streamwise and the cross-stream direction, we used a two-dimensional boundary element method (BEM) based on potential flow theory in which the flow is assumed to be irrotational, incompressible and inviscid. Previously, this method was used to model flow over unsteady hydrofoils \cite{katz2001low, Moored2018Bem, moored2019inviscid}, and their interaction with a solid boundary and the associated performance for constrained and unconstrained foils \cite{quinn2014unsteady, kurt2019swimming}. Further details about the numerical method used here can be found in previous work \cite{moored2019inviscid, kurt2019swimming}. 

Like the constrained experiments, the freely-swimming simulated foils were prescribed sinusoidal pitching motions about their leading edge with an out-of-phase synchrony ($\phi=\pi$). Each foil is assigned with a mass normalized by their characteristic added mass, $m^* = m/(\rho sc^2 ) = 2.76$ and $1.74$ in stream-wise and cross-stream-wise directions, respectively, similar to the \textit{unconstrained foil experiments}, which is comparable with biology (for example, $m^* = 3.86$ was calculated for cod in \cite{Akoz2018}).  Note that the constrained foil measurements presented earlier have an effective dimensionless mass of $m^*=\infty$, since the foils do not exhibit recoil motions. In the simulations, since the hydrofoils are self-propelled a drag force is applied to them, which follows a high Reynolds number $U^2$ drag law \cite{Munson1998}. The drag force is determined by the properties of a virtual body, not present in the computational domain, which is determined by the Lighthill number, $Li$.  This dimensionless number is equal to $Li = C_D S_{wp}$, where $C_D$ is the drag coefficient of the virtual body and $S_{wp}$ is a ratio of the wetted area of the virtual body and propulsor to the planform area of the propulsor.  This characterizes how the virtual body drag coefficient and the body-to-propulsor sizing affect the balance of the thrust and drag forces. As previously shown in \cite{moored2019inviscid}, when $Li$ decreases at a fixed frequency and amplitude of motion,  free-swimming speed increases. This leads to a decrease in $St$ and, consequently, a decrease in the equilibrium distance as reported in \cite{kurt2019swimming}. In steady-state free-swimming the $Li$ is equal to the dynamic pressure-based thrust coefficient, that is, $Li = C_T$.  To investigate the stability of the side-by-side equilibrium with recoil motions present, the Lighthill number was then set to $Li = 0.3$. This was the lowest value achievable for the numerical stability of the current BEM formulation.

} 

\showmatmethods

 % Display the Materials and Methods section

%%%%%%%%%%%%%%%%%%%%%%%%%%%%

%%%%%%%%%%%%%%%%%%%%%%%%%%%%

\acknow{This work was supported by the National Science Foundation under Program Director Dr. Ronald Joslin in Fluid Dynamics within CBET on NSF award number 1653181 and NSF collaboration award number 1921809. Some of this work was also funded by the Office of Naval Research under Program Director Dr. Robert Brizzolara on MURI grant number N00014-08-1-0642 and N00014-22-1-2616.}
\showacknow{} % Display the acknowledgments section

% Bibliography
\bibliography{pnas-template-main}

\end{document}